\documentclass{aa}
 \usepackage{amssymb}
\usepackage{latexsym}
\usepackage{url}
\usepackage{xcolor}
\definecolor{newcolor}{rgb}{.8,.349,.1}
\usepackage{amsmath}
\usepackage{threeparttable}
\usepackage{cleveref}
\usepackage{graphicx,caption}

\usepackage[varg]{txfonts}
 \usepackage[]{natbib}

\begin{document}

\title{Potential magnetic field calculator for solar physics applications using staggered grids} 
\author{Callum M. Boocock\inst{1}
  \and David Tsiklauri\inst{1} }
\titlerunning{Potential field calculator for staggered magnetic grids}
\institute{School of Physics and Astronomy, Queen Mary University of London, G.O. Jones Building, 327 Mile End Road, London, E1 4NS, UK} 
\date{Received 19th November 2018 / Accepted 19th March 2019}
\abstract{A program has been designed to generate accurately a potential magnetic field on a staggered grid by extrapolating the magnetic field normal to the photospheric surface. The code first calculates a magnetic potential using the Green's function method and then uses a finite differencing scheme to calculate the magnetic field from the potential. A new finite differencing formula was derived which accounts for  grid staggering; it is shown that this formula gives a numerical approximation that is closest to the real potential field. It is also shown that extending the region over which normal photospheric field is specified  can improve the accuracy of the potential field produced. The program is a FORTRAN 90 code that can be used to generate potential magnetic field inputs for Lare3d and other MHD solvers that use a staggered grid for magnetic field components. The program can be parallelised to run quickly over multiple computing cores. The code and supporting description are provided in the appendices and at \protect\url{https://github.com/calboo/Potential\_Field\_Calculator}.} 
\maketitle 

\section{Introduction}
\label{section1}
The coronal heating problem, the question of why the Sun's corona is much hotter ($\sim$1 MK) than photosphere ($\sim$6000 K), is an ongoing problem in solar physics. Researchers agree that energy is transported to the corona by non-thermal transport of energy through the Sun's magnetic field \citep{Arregui}, although the dominant mechanism for coronal heating is under debate \citep{Parnell}. Computational 3D magnetohydrodynamic (MHD) models are often used to address the coronal heating problem \citep{klimchuk}.

Computational MHD models require an initial magnetic field to be specified. For many problems, primarily those concerning MHD waves, it is useful to have a static equilibrium on which to impose perturbations to the velocity and magnetic fields \citep{goosens}. In the case of a static equilibrium both the velocity vector $\mathbf{v}$ and its time derivative $\partial\mathbf{v}/\partial t$ must equal zero. 

%
%

Analytic expressions exist for a number of structures that describe magnetostatic equilibria in both 2D and 3D for example \citep{Smith,RudermanA,Oliver,Cuperman,Gent}. For more detailed magnetic structures we need information from solar observations. Maps of the line-of-sight and vector magnetograms from the photosphere can be measured by means of spectropolarimetric methods such as the Zeeman effect, i.e. the splitting of spectral lines in the presence of a magnetic field \citep{Beckers}. It is however much more difficult to measure directly the magnetic field in the solar corona \citep{Ruan}. Fortunately the photospheric magnetic field can be used to reconstruct the coronal field by means of extrapolation. Extrapolation of the photospheric magnetic field is currently the primary tool for modelling the coronal magnetic fields \citep{Tadesse}.

There are many methods for extrapolating the structure of the magnetic field from surface measurements. The commonly used methods rely on various assumptions \citep{Neukirch}. Most commonly non-magnetic forces such as pressure gradients and gravity are neglected; this is well justified  in the solar corona because of the low plasma beta \citep{WiegFF} and when considering scales smaller than the hydrostatic scale height \citep{Peter}. For magnetostatic equilibrium the magnetic Lorentz force must then be zero $\mathbf{J}\times\mathbf{B} = 0$. This defines a force-free field \citep{WiegSak}. Another way of expressing this condition is given below in \cref{eq:2} where $\alpha$ can be either zero, a constant, or a variable that is constant along field lines. These cases correspond to potential, linear, and non-linear force-free fields, respectively \citep{Ashwan}. This expression is written as

\begin{equation}\label{eq:2}
\nabla\times\mathbf{B} = \alpha\mathbf{B}.\end{equation}

Various methods of extrapolation are used to reconstruct potential fields, linear force-free fields \citep{Gary}, and non-linear force-free fields \citep{WiegFF}. There are also methods of extrapolation available to reconstruct non-force-free fields although these may require additional data \citep{Wieg2004}. Furthermore several of these methods can be extended from Cartesian to spherical coordinates so that the solar coronal field can be calculated globally \citep{Wieg2007,WN2008}.

In the particular case of a potential force-free field there is zero current $\mathbf{J} = \nabla\times\mathbf{B} = 0$. The field is called a potential magnetic field because the magnetic field can be expressed in terms of a scalar potential $\mathbf{B}  = \nabla\phi$. Potential magnetic fields are useful because they can be used, for example as an initial topology for MHD simulations \citep{Gudiksen, Masson, Bingert1, Bingert2, Bourdin}, to study MHD wave phenomena \citep{Thackray,Ofman,Smith,RudermanA}, to embed non-potential fields \citep{Browning}, or to generate non-force-free MHD equilibria \citep{Gordov,Solanki,Khomenko,Pizzo,WN2006,InoueRev,Inoue}.

To calculate a potential magnetic field from the normal magnetic field $B_n$ at the photosphere many methods can be utilised such as the Fourier expansion, spherical harmonic expansions, and Green's function. The Fourier method and the Green's function method are compared in \cite{sakurai}. Extrapolated fields require that the boundary conditions for extrapolation are able to match those of the MHD simulation \citep{Otto}. 

In this work we consider the Green's function method, which uses a discrete approximation for $B_n$ within each mesh. This method is better suited for considering non-periodic, isolated magnetic regions but has the drawback of assuming that $B_n = 0$ outside the region of interest. This method was first used by Schmidt \citep{Schmidt}, was later developed for oblique photospheric data by Semel \citep{Semel}, and adapted further by Sakurai who adjusted the method for practical applications on a finite numerical grid \citep{sakurai}.

This paper describes a potential field calculator based on a modified Green's function method. The purpose of this paper is to demonstrate an improved method of finite differencing suited to a MHD solver using a staggered magnetic field. It also demonstrates that by increasing the extent of input data relative to the computational domain the errors associated with assuming $B_n = 0$ outside the region of interest can be reduced.

The different methods of numerical differentiation that were considered for the potential field calculator are explained and compared with the most accurate methods being chosen for the final implementation. The calculator has been designed to produce initial magnetic field inputs for Lare3d \citep{lare} but works with any MHD solver that uses staggered grids for the magnetic field components. 

\section{Green's function method}
\label{section3}
Taking $\mathbf{J} = \nabla\times\mathbf{B} = 0$ as an initial condition, the Gauss law of magnetism can be expressed as the Laplace equation for the magnetic potential \citep{Ashwan}.
\begin{equation}\label{eq:3}
\nabla\cdot\mathbf{B}=\nabla^2\phi = 0,
\end{equation}

By solving the Laplace equation for the scalar potential $\phi$,  a potential magnetic field $\mathbf{B}$ can be calculated that results in a static equilibrium. Consider a 3D domain with $z$ vertical and the photospheric boundary at $z=0$. We denote the magnetic field normal to the photospheric boundary as $B_z(x,y,z),$ and the normal field at the photospheric boundary is then $B_{z0}(x,y)=B_z(x,y,0)$. 

The Laplace equation with the photospheric boundary condition $B_{z0}=\partial\phi/\partial z$ can be solved numerically on a discrete grid of points (i,j,k) with grid separation $d$ by using the modified Green's function method given below in \cref{eq:4} \citep{sakurai}, i.e.

\begin{equation}\label{eq:4}
\phi(i,j,k) = \sum_{i_1}\sum_{j_1} \frac{B_{z0}(i_1,j_1)d}{2\pi\sqrt{(i-i_1)^2+(j-j_1)^2+(k+\frac{1}{\sqrt{2\pi}})^2}},
\end{equation}

where $i_1$ and $j_1$ are dummy variables used to sum over contributions from each point at the photosphere. The normal magnetic field contributes to the potential via a discrete approximation of the Green's function. By virtue of the chosen Green's function the potential satisfies the Laplace equation and boundary conditions at $z=0$ and at $r\to\infty$ as follows:

\begin{equation}\label{eq:5a}
\begin{split}
-\mathbf{n}\cdot\nabla\phi = B_{z0} \qquad (z=0) \\
\lim_{r \to \infty} \phi(\mathbf{r}) = 0 \qquad (z>0)
\end{split}
\end{equation}

This particular choice of Green's function is representative of a ``submerged magnetic source'' at a depth $d/\sqrt{2\pi}$ beneath each grid point at $z=0$. There is a slight inconvenience with this choice of Green's function in that the calculated normal magnetic field does not exactly match the original normal field at $z=0$. Whilst there are other possible choices that give a closer match, one of which is described in \citep{sakurai}, this particular Green's function has been chosen for computational convenience because the focus of this paper is not on the choice of Green's function but rather the method of finite differencing used to differentiate the magnetic potential.

Once the magnetic potential is determined at each grid point the magnetic field components can then be calculated as derivatives of the magnetic potential,

\begin{equation}\label{eq:5}
B_x = \frac{\partial \phi}{\partial x},\qquad B_y = \frac{\partial \phi}{\partial y},\qquad B_z =  \frac{\partial \phi}{\partial z}.\end{equation}

If a finite differencing scheme is used to differentiate the magnetic potential then the potential needs to be calculated in additional ``ghost'' cells at the boundaries of the domain. The number of additional cells depends upon the order of the differencing scheme. Ghost cells are also required in the input for Lare3d as described in \cref{section2}. 

As the potential must be calculated in the ghost cells, the potential due to the magnetic source, represented by the Green's function, must also be positioned beneath these cells and outside of the domain. This can be done by either increasing the indices in the z direction so that the ghost cells are raised above the magnetic source or by altering \cref{eq:4} to effectively lower the magnetic source by the required number of cells.

In either case the overall effect is to consider the lowest ghost cell as the new position of the initial normal field. The magnetic field considered in the MHD solver is then be shifted up from the photosphere by the length of the additional ghost cells. Whilst this inaccuracy in position is inconvenient and there are ways to avoid this, for example manually fixing boundary conditions in Lare3d or not differentiating to determine the magnetic field in the lowest two layers, we decided to make this modification on the basis that any manual fixing of the boundary conditions affects the potential field equilibrium in ways that are independent of the methods of the finite differencing under investigation. The modified Green's function method used in this paper is therefore a slight modification of \cref{eq:4}, which submerges the magnetic source represented by the Green's function by an additional $n$ cell depths, where $n$ is the maximum number of ghost cells required, i.e.

\begin{equation}\label{eq:8}
\phi(i,j,k) = \sum_{i_1}\sum_{j_1} \frac{B_{z0}(i_1,j_1)d}{2\pi\sqrt{(i-i_1)^2+(j-j_1)^2+(k+n+\frac{1}{\sqrt{2\pi}})^2}}.\end{equation}

\section{Staggered grids}
\label{section2}
\begin{figure}
        \resizebox{\hsize}{!}{\includegraphics{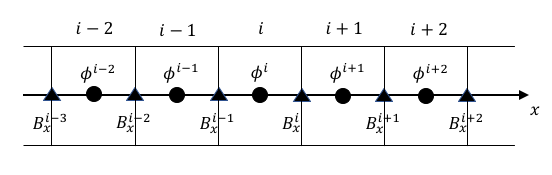}}
        \caption{One-dimensional representation of a small section around the $i^{th}$ cell in the $x$ direction showing the position of $\phi$ values at the cell 
        centres and $B_x$ values at the cell boundaries in the $x$ direction.}
        \label{fig:gridstag}
\end{figure}

The computational domain in Lare3d is of size $(nx,ny,nz)$ however the magnetic field is defined on a staggered grid so that the magnetic field components are collocated at the face-centres of each cell. For example, the values of $B_x$ are shifted away from the cell centre by half a cell in the $x$ direction, as shown in \cref{fig:gridstag}, and similarly for $B_y$ and $B_z$ in the $y$ and $z$ directions, respectively. The magnetic field components are staggered in this way to maintain the solenoidal property of the magnetic field $\nabla\cdot\mathbf{B}=0$ as it is evolved via the induction equation \citep{balsara}. The grid staggering changes the position at which the magnetic field components must be specified in relation to the magnetic potential. Owing to the staggered grid the extents of $B_x$, $B_y$ and $B_z$ within the computational domain are written as

\begin{equation}\label{eq:6}
\begin{split}
B_x(0:nx,1:ny,1:nz), \\
B_y(1:nx,0:ny,1:nz), \\
B_z(1:nx,1:ny,0:nz).
\end{split}
\end{equation}

The numerical grid on which the initial magnetic field must be specified in Lare3d is however larger. This is because Lare3d is second order accurate in space and therefore requires two ghost cells at each boundary. The extents of $B_x$, $B_y$ and $B_z$ to be specified in the initial magnetic field are therefore written as

\begin{equation}\label{eq:7}
\begin{split}
B_x(-2:nx+2,-1:ny+2,-1:nz+2), \\
B_y(-1:nx+2,-2:ny+2,-1:nz+2), \\
B_z(-1:nx+2,-1:ny+2,-2:nz+2).
\end{split}
\end{equation}
\section{Comparison methodology}
\label{section4}

\begin{figure*}[h!]
\begin{minipage}{\columnwidth}
        \resizebox{\hsize}{!}{\includegraphics{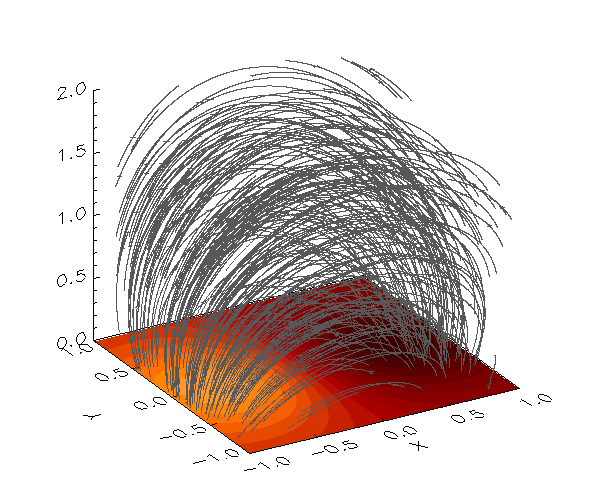}}
        \caption{Field line diagram of the magnetic field used for Test Case 1. The photospheric surface at the base of the domain is a coloured contour 
        of $B_{z0}$.}
        \label{fig:test1}
\end{minipage}\hfill
\begin{minipage}{\columnwidth}
        \resizebox{\hsize}{!}{\includegraphics{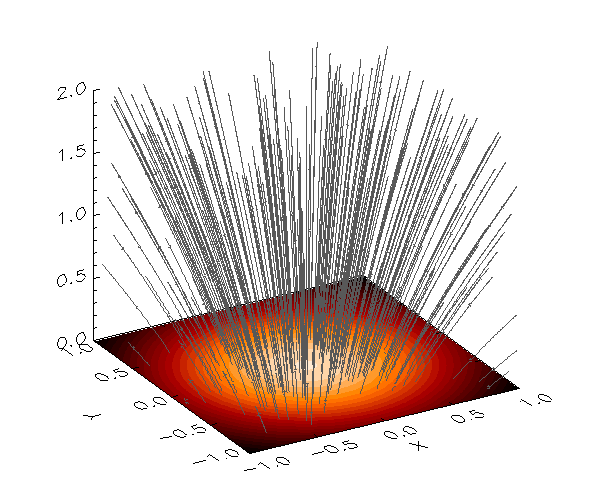}}
        \caption{Field line diagram of the magnetic field used for Test Case 2. The photospheric surface at the base of the domain is a coloured contour of $B_{z0}$.}
        \label{fig:test2}
\end{minipage}\hfill
\end{figure*}

In order to compare the effects of different methods of numerical differentiation and the extent of the region over which the normal photospheric field is specified, we need to produce and compare different potential fields. Our aim is to maximise the accuracy of the potential field calculator. A potential magnetic field should be current free, should not produce motion in a static plasma, and should not evolve with time. 

\subsection{Test cases}

For this study we used two analytically defined test cases. The first test case is a region containing a bipole loop and the second is a region containing a unipolar magnetic field. For both test cases magnetic fields are calculated over a domain size of $nx=ny=nz=200$ with grid spacing of $d = 0.01$. Defining the extends of our domain as [-1:1,-1:1,0:2]  with its origin at (0,0,0), the test cases can be defined in terms of the input magnetic field $B_{z0}$  as follows: \\

\textbf{Test Case 1} \\

The formula for a bipole loop is determined by taking the analytic expression for a force-free bipole field given in \citep{Cuperman}. We take the current free version of these equations and set the coordinates of the magnetic poles to $(x_{01},y_{01},z_{01}) = (-0.5,0,-1)$ and $(x_{02},y_{02},z_{02}) = (0.5,0,-1)$. Finally we take the $B_z$ component and set $z=0$ to arrive at 
\begin{equation}\label{eq:9}
B_{z0} = A\left[\left(\frac{1}{((x+0.5)^2+y^2+1)^{3/2}}\right)-\left(\frac{1}{((x-0.5)^2+y^2+1)^{3/2}}\right)\right].
\end{equation} 

\textbf{Test Case 2} \\

The formula for a unipolar magnetic field is determined by assuming that the strength of the normal magnetic field $B_z$ drops off as a Gaussian with distance from a magnetic source term. We take the position of the magnetic source term to be at the origin a unit distance beneath the photosphere to arrive at
\begin{equation}\label{eq:10}
B_{z0} = A \exp\left(-\frac{x^2+y^2+1}{2}\right),
\end{equation} 

The potential fields produced by these test cases are represented by field line diagrams in \cref{fig:test1,fig:test2}; the shading at the base of the domains in these images represents the strength and direction of the magnetic field at the photosphere.
\subsection{Metrics}
To analyse the magnetic fields that the potential field calculator produces we import each field into the MHD solver Lare3d and measure the current across the domain. 

We then set the initial velocity to zero, the initial pressure to a constant value of $P=5\cdot 10^{-9}$, and resistivity to $\eta=5\cdot 10^{-5}$. We set $\gamma = 5/3$ so the fluid is adiabatic and set the shock viscosities to $\nu_1=0.1$ and $\nu_2=0.5$. We set the boundary conditions such that the velocity and magnetic field have fixed values at the boundaries. 

We then allow the simulation to evolve for  $100\tau_A$ where $\tau_A$ is the Alfv\'en time. The number of timesteps this requires varies as Lare3d uses a CFL limited timestep at each iteration. This is not intended as a relaxation phase but rather test of how stable the potential field configuration is over time. Finally we measure the current, velocity, and difference between the final and initial magnetic fields across the domain.

The metrics used to analyse the effectiveness of each method are given below. The closer these values are to zero the closer the field is to a potential field and the closer our equilibrium is to static equilibrium. The average current densities are the mean value of the current density $\mathbf{J}$ over all grid points.
These metrics include\begin{tabbing}
MAJ$_0$ \qquad \= The maximum absolute current density at time zero. \\
J$_{avg0}$ \> The average absolute  current density at time zero. \\
MAJ$_f$ \> The maximum absolute  current density after $100\tau_A$. \\
J$_{avgf}$ \> The average absolute  current density after $100\tau_A$. \\
MAV \> The maximum absolute velocity after $100\tau_A$. \\
$\Delta B_{max}$ \> The maximum difference between \\
\> initial and final magnetic fields. 
\end{tabbing}
\section{Differentiation of magnetic potential $\phi$}
\subsection{Method}
\label{section5}
In order to calculate the magnetic field components in our domain we need to differentiate the magnetic field potential according to \cref{eq:5}. The effect of grid staggering is that the magnetic field component in a particular direction must be calculated at the cell edges in that direction not the cell centre. The magnetic potential $\phi$ however is calculated at the cell centres. This is illustrated in \cref{fig:gridstag}. 

In this section three methods of numerical differentiation are presented, which each account differently for this discrepancy. The calculation of $B_x(i,j,k)$ is taken as an example. The indices $j$ and $k$ are taken to be fixed and $B_x(i,j,k)$ is denoted simply as $B^{i}_x$. \\ 

\textbf{Method A} - In this method a central difference formula of order $\mathcal{O}(h^4)$ is used and the grid staggering is not taken into account. The magnetic field calculated using method A is denoted as $B^{i}_{xA}$. The formula used to calculate $B^{i}_{xA}$ is given in \cref{eq:11} below. This method requires an additional four ghost cells in each direction, i.e.

\begin{equation}\label{eq:11}
B^{i}_{xA}= \frac{\phi^{i-2}-8\phi^{i-1}+8\phi^{i+1}-\phi^{i+2}}{12d}
.\end{equation} \\

\textbf{Method B} - In this method the grid staggering is accounted for by calculating the magnetic field components at the cell centres using method A and then averaging these values to get the magnetic field components at the cell edges. The magnetic field calculated using method B is denoted as $B^{i}_{xB}$. The formula used to calculate $B^{i}_{xB}$ is given in \cref{eq:12} below. This method requires an additional five ghost cells in each direction, i.e.

\begin{equation}\label{eq:12}
B^{i}_{xB} = \frac{B^{i}_{xA} + B^{i+1}_{xA}}{2}
.\end{equation} \\

\textbf{Method C} - In this method the grid staggering is accounted for by calculating the magnetic field components using a different central difference formula. A full derivation of this formula is given in \cref{App1}. The magnetic field calculated using method C is denoted as $B^{i}_{xC}$. The formula used to calculate $B^{i}_{xC}$ is given in \cref{eq:13} below. This method requires an additional three ghost cells in each direction, i.e.

\begin{equation}\label{eq:13}
B^{i}_{xC}= \frac{\phi^{i-1}-27\phi^{i}+27\phi^{i+1}-\phi^{i+2}}{24d}
.\end{equation} 
\subsection{Results}
For each method of differentiation we generated a potential field for each test case. The potential fields were generated from input fields $B_{z0}$ defined over the base of our computational domain. We analysed and compared the potential fields generated and the results of our analysis are given in \cref{tab:1,tab:2}. 

For Test Case 1 the potential field produced using Method 1 was not run to 100$\tau_A$. The currents present in the initial magnetic field caused $\nabla\cdot\mathbf{v}$ to increase, significantly reducing $\rho$ in some parts of the domain. This led to a significant reduction in the CFL limited timestep as the simulation progressed making it take unfeasibly long to run to a simulation time of 100$\tau_A$; the values given are instead for 50$\tau_A$. 

From \cref{tab:1,tab:2} we can see that the magnetic field generated is closest to a potential field when Method C is used.  Although the difference between using Methods B and C can be marginal Method C is the most accurate method of numerical differentiation. Furthermore Method C has the advantage of being more computationally efficient, as it does not require any averaging after the initial differentiation and requires the least number of additional ghost cells for the potential calculation. Method C was therefore selected as the preferred option for our potential calculator.

\begin{table}
\caption{Value of each metric for Method A, Method B, and Method C of numerical differentiation for Test Case 1.}
\label{tab:1}
\centering
\begin{tabular}{ |p{0.2\columnwidth}||p{0.2\columnwidth}|p{0.2\columnwidth}|p{0.2\columnwidth}|  }
 \hline
 \multicolumn{4}{|c|}{Test Case 1} \\
 \hline
  & Method A$^*$ &Method B & Method C\\
 \hline
 MAJ$_0$                & $4.45 \cdot 10^{-1}$  & $1.72 \cdot 10^{-1}$  & $7.86 \cdot 10^{-3}$\\
 J$_{avg0}$             & $2.81 \cdot 10^{-3}$  & $3.23 \cdot 10^{-5}$   & $1.61 \cdot 10^{-5}$\\
 MAJ$_f$                & $3.79 \cdot 10^{-1}$  & $2.01 \cdot 10^{-2}$  & $1.97 \cdot 10^{-2}$\\
 J$_{avgf}$             & $2.58 \cdot 10^{-3}$  & $1.16 \cdot 10^{-4}$  & $1.01 \cdot 10^{-4}$\\
 MAV                    & $1.14 \cdot 10^{-2}$  & $2.27 \cdot 10^{-5}$  & $1.83 \cdot 10^{-5}$\\
 $\Delta B_{max}$       & $8.14 \cdot 10^{-3}$  & $7.02 \cdot 10^{-4}$  & $4.06 \cdot 10^{-4}$\\
 \hline
\end{tabular}
\end{table}

\begin{table}
\caption{Value of each metric for Method A, Method B, and Method C of numerical differentiation for Test Case 2.}
\label{tab:2}
\centering
\begin{tabular}{ |p{0.2\columnwidth}||p{0.2\columnwidth}|p{0.2\columnwidth}|p{0.2\columnwidth}|  }
 \hline
 \multicolumn{4}{|c|}{Test Case 2} \\
 \hline
 & Method A &Method B & Method C\\
 \hline
 MAJ$_0$                & $2.97 \cdot 10^{-1}$  & $1.15 \cdot 10^{-2}$  & $5.25 \cdot 10^{-3}$\\
 J$_{avg0}$             & $1.97 \cdot 10^{-3}$  & $2.98 \cdot 10^{-5}$   & $1.48 \cdot 10^{-5}$\\
 MAJ$_f$                & $2.13 \cdot 10^{-1}$  & $1.40 \cdot 10^{-2}$  & $1.37 \cdot 10^{-2}$\\
 J$_{avgf}$             & $2.10 \cdot 10^{-3}$  & $7.58 \cdot 10^{-5}$  & $7.13 \cdot 10^{-5}$\\
 MAV                    & $3.38 \cdot 10^{-3}$  & $1.36 \cdot 10^{-5}$  & $1.12 \cdot 10^{-5}$\\
 $\Delta B_{max}$       & $2.40 \cdot 10^{-3}$  & $2.82 \cdot 10^{-4}$  & $2.10 \cdot 10^{-4}$\\
 \hline
\end{tabular}
\begin{tablenotes}
\footnotesize
\item $^*$For Test Case 1, Method A the values are for 50$\tau_A$ instead of 100$\tau_A$.
\end{tablenotes}
\end{table}

\begin{figure*}[h!]
\begin{minipage}{\columnwidth}
        \resizebox{\hsize}{!}{\includegraphics{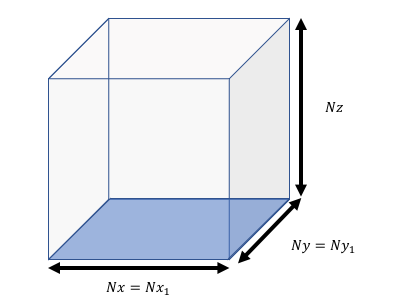}}
        \caption{llustration of the computational domain for $\phi$, shown as the grey box, and the input region for $B_{z0}$, shown as the blue square, when $(Nx_1,Ny_2)$ is equal to $(Nx,Ny)$.}
        \label{fig:domain1}
\end{minipage}\hfill
\begin{minipage}{\columnwidth}
        \resizebox{\hsize}{!}{\includegraphics{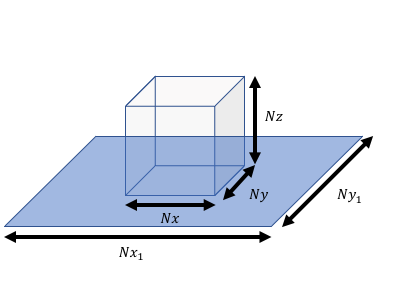}}
        \caption{Illustration of the computational domain for $\phi$, shown as the grey box, and the input region for $B_{z0}$, shown as the blue square, when $(Nx_1,Ny_2)$ is equal to $(3Nx,3Ny)$.}
        \label{fig:domain2}
\end{minipage}\hfill
\end{figure*}

\section{Size of input region for $B_{z0}$}
\subsection{Method}
\label{section6}

We now consider the size of the input region, that is the area at the photospheric boundary over which the input field $B_{z0}$ is specified. The size of the computational domain for $\phi$ is denoted as $(Nx,Ny,Nz)$ and the size of the input region for $B_{z0}$ is denoted as $(Nx_1,Ny_2)$. So far the input field $B_{z0}$ has only been defined over the base of our computational domain. In other words $(Nx_1,Ny_2)$ has been set to exactly equal $(Nx,Ny)$. The input field $B_{z0}$ can however be defined over a larger area. 

\begin{figure*}[h!]
\begin{minipage}{\columnwidth}
        \resizebox{\hsize}{!}{\includegraphics{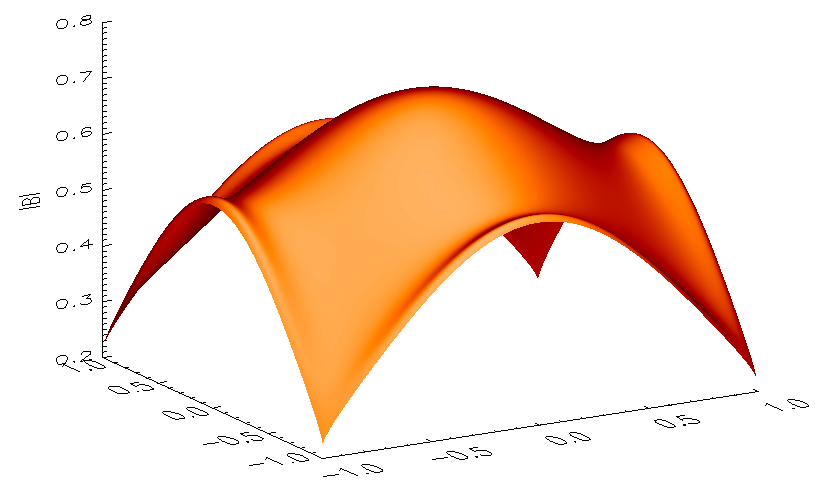}}
        \caption{Shaded surface diagram of the magnetic field strength at $z=0$ for the potential field generated using the standard input region. 
        The flared edges are due to the discontinuity at the edge of the input region.}
        \label{fig:flared}
\end{minipage}\hfill
\begin{minipage}{\columnwidth}
        \resizebox{\hsize}{!}{\includegraphics{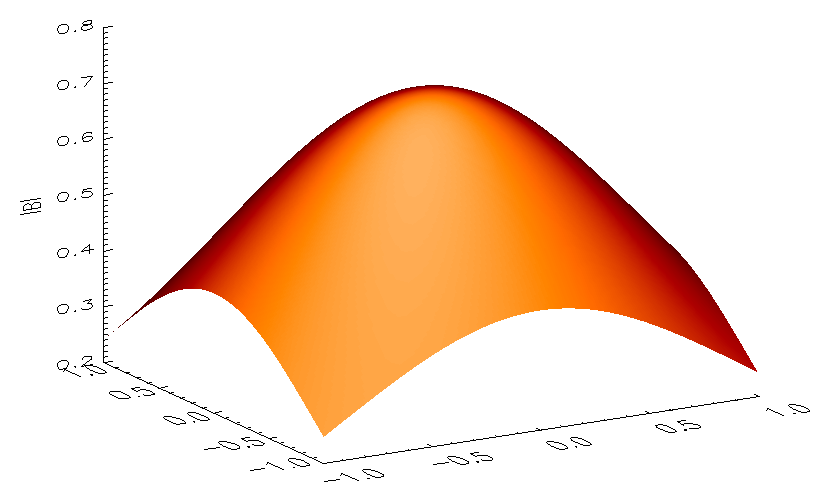}}
        \caption{Shaded surface diagram of the magnetic field strength at $z=0$ for the potential field generated using the extended input region.
        The edges are smoother because of the extended input region.}
        \label{fig:smooth}
\end{minipage}\hfill
\end{figure*}

The motivation for extending our input region is that the current $J$ in our previous potential fields has been concentrated at the $x/y$ boundaries. The reason for this can be seen in \cref{fig:flared}; the gradient of the magnetic field strength changes suddenly at these boundaries. By extending the input region we attempt to reduce the effect of this change in gradient and reduce the current at the boundaries. We consider two options for the size of the input region: \\

\textbf{Standard input region} - $(Nx_1,Ny_2)$ is set to equal exactly  $(Nx,Ny)$ and the centres of the input region and computational domain are aligned. The input data then covers exactly the same area of photosphere as the computational domain for $\phi$. This is illustrated in \cref{fig:domain1}. \\

\textbf{Extended input region} - $(Nx_1,Ny_2)$ is set equal to $(3Nx,3Ny)$ and the centres of the input region and computational domain are aligned. The input data then covers a much larger area of photosphere than the computational domain for $\phi$. This is illustrated in \cref{fig:domain2}.

\subsection{Results}

For each test case we produce potential fields using both the standard and extended input regions. We used \cref{eq:8} to calculate our magnetic potential $\phi$ and Method C to perform our numerical differentiation. The potential fields generated were each run in Lare3d for 100$\tau_A$ before being analysed and compared. The results of our analysis are given in \cref{tab:3,tab:4}. We can see that the magnetic field generated using a larger input region is closer to a potential field. Furthermore we can see from \cref{fig:flared,fig:smooth} that using a larger input region produces magnetic fields that are smoother at the boundaries. 

It is worth noting that because of the nature of the Green's function method used the potential calculator expects $B_{z0}$ to equal zero outside of the input region. If significant magnetic fields are expected outside of the input region then the accuracy of the calculator can be improved by  extending the size of the input region; however, there is a trade-off between accuracy and computing time, besides which limited photospheric data may be available.

\begin{table}[!h]
\caption{Table showing the value of each metric using standard and extended input regions for Test Case 1.}
\label{tab:3}
\centering
\begin{tabular}{ |p{0.2\columnwidth}||p{0.3\columnwidth}|p{0.3\columnwidth}| }
 \hline
 \multicolumn{3}{|c|}{Test Case 1} \\
 \hline
 & Standard & Extended\\
 \hline
 MAJ$_0$                & $7.86 \cdot 10^{-3}$ & $1.01 \cdot 10^{-4}$\\
 J$_{avg0}$             & $1.61 \cdot 10^{-5}$ & $6.42 \cdot 10^{-6}$ \\
 MAJ$_f$                & $1.97 \cdot 10^{-2}$ & $1.90 \cdot 10^{-3}$\\
 J$_{avgf}$             & $1.01 \cdot 10^{-4}$ & $5.35 \cdot 10^{-5}$\\
 MAV                    & $1.83 \cdot 10^{-5}$ & $1.74 \cdot 10^{-5}$\\
 $\Delta B_{max}$       & $4.06 \cdot 10^{-4}$ & $1.64 \cdot 10^{-4}$\\
 \hline
\end{tabular}
\end{table}

\begin{table}[!h]
\caption{Table showing the value of each metric using standard and extended input regions for Test Case 2.}
\label{tab:4}
\centering
\begin{tabular}{ |p{0.2\columnwidth}||p{0.3\columnwidth}|p{0.3\columnwidth}| }
 \hline
 \multicolumn{3}{|c|}{Test Case 2} \\
 \hline
 & Standard & Extended\\
 \hline
 MAJ$_0$                & $5.25 \cdot 10^{-3}$& $7.99 \cdot 10^{-6}$\\
 J$_{avg0}$             & $1.48 \cdot 10^{-5}$& $1.72 \cdot 10^{-6}$\\
 MAJ$_f$                & $1.37 \cdot 10^{-2}$& $5.64 \cdot 10^{-4}$\\
 J$_{avgf}$             & $7.13 \cdot 10^{-5}$& $1.39 \cdot 10^{-5}$\\
 MAV                    & $1.12 \cdot 10^{-5}$& $2.55 \cdot 10^{-7}$\\
 $\Delta B_{max}$       & $2.10 \cdot 10^{-4}$& $2.40 \cdot 10^{-5}$\\
 \hline
\end{tabular}
\end{table}
\section{Comparison with analytical bipole loop}
\subsection{Method}
\label{section7}
In \cref{section5} we determined the optimal method of numerical differentiation (Method C) and in \cref{section6} we saw that using an extended input region produces fields that are closer to a potential field solution. We now compare a potential field solution calculated with these improvements to an analytical solution for the same potential field.

Using our chosen method of differentiation and extended input region we apply our Green's method for extrapolation, as detailed in \cref{section2}, to calculate a potential field. This field is produced using the normal photospheric field for a bipole loop, given as Test Case 1 in \cref{section4}. We compare this field solution to the analytic solution for a potential bipole field given in \cite{Cuperman}, which has the same normal magnetic field at the photosphere. This is given by
\begin{equation}\label{eq:14}
\begin{split}
&B_x = B_{x,1} - B_{x,2}, \quad
B_y = B_{y,1} - B_{y,2}, \quad
B_z = B_{z,1} - B_{z,2}, \\ 
&B_{x,j} = \frac{x_j}{R_j^3}, \quad
B_{y,j} = \frac{y_j}{R_j^3}, \quad
B_{z,j} = \frac{z_j}{R_j^3}, \\ 
&R_j^2 = x_j^2 +y_j^2 + z_j^2, \\
&x_j = x - x_{0,j}, \quad
y_j = y - y_{0,j}, \quad
z_j = z - z_{0,j}
\end{split}
,\end{equation}
where $R_j$ are the distances from each magnetic pole for $j=1,2$ positioned at $(x_{01},y_{01},z_{01}) = (-0.5,0,-1)$ and $(x_{02},y_{02},z_{02}) = (0.5,0,-1)$.
\subsection{Results}
The magnetic field produced by our potential calculator and the potential field given by \cref{eq:14} were both run in Lare3d and analysed in the same way as for the previous magnetic fields. This was done to compare both the initial magnetic fields and the evolution of these field after $100\tau_A$ of simulation time in Lare3d.

The results of this analysis are given in \cref{tab:5}. The currents present in the analytical field can be explained by numerical diffusion. Numerical diffusion is caused by the central differencing scheme used to calculate the currents in Lare3d. The truncation error of the central difference scheme used to calculate the currents is of the order $\mathcal{O}(h^2)$, where h is the grid spacing. Indeed we can see that maximum current in this solution is also of the order $\mathcal{O}(h^2)$ with grid spacing $h=0.01$.

\begin{table}[h!]
\caption{Value of each metric for the analytical and numerically calculated potential fields used for Test Case 1.}
\label{tab:5}
\centering
\begin{tabular}{ |p{0.2\columnwidth}||p{0.3\columnwidth}|p{0.3\columnwidth}| }
 \hline
 & Analytical & Numerical \\
 \hline
 MAJ$_0$                & $1.20 \cdot 10^{-4}$ & $1.01 \cdot 10^{-4}$\\
 J$_{avg0}$             & $7.54 \cdot 10^{-6}$ & $6.42 \cdot 10^{-6}$ \\
 MAJ$_f$                & $2.18 \cdot 10^{-3}$ & $1.90 \cdot 10^{-3}$\\
 J$_{avgf}$             & $6.14 \cdot 10^{-5}$ & $5.35 \cdot 10^{-5}$\\
 MAV                    & $1.70 \cdot 10^{-5}$ & $1.74 \cdot 10^{-5}$\\
 $\Delta B_{max}$       & $2.32 \cdot 10^{-4}$ & $1.64 \cdot 10^{-4}$\\
 \hline
\end{tabular}
\end{table}

These results show that the currents in the numerical solution as calculated by the potential calculator are similar and slightly smaller than those for the analytical solution. We can therefore say that, for this example, the numerical field solution of our potential calculator is correct to within the numerical diffusion of the simulation grid.

\begin{figure*}[!]
\begin{minipage}{\columnwidth}
        \resizebox{\hsize}{!}{\includegraphics{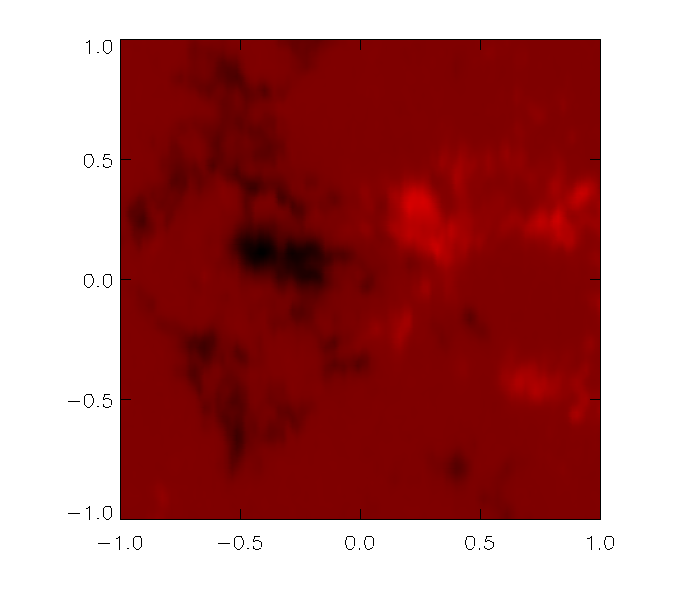}}
        \caption{Contour plot of the original magnetogram data of the normal magnetic field at the photosphere. The colour indicates the direction and 
        strength of the magnetic field.}
        \label{fig:bz0}
\end{minipage}\hfill
\begin{minipage}{\columnwidth}
        \resizebox{\hsize}{!}{  \includegraphics{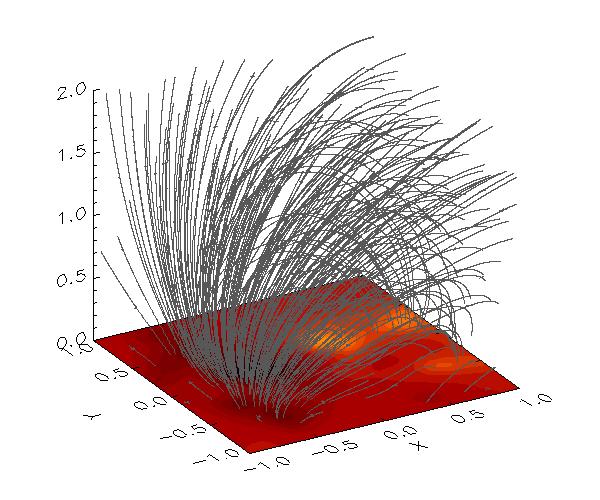}}
        \caption{Field line diagram of the potential magnetic field produced using solar magnetogram data. The photospheric surface at the base of the 
        domain is a coloured contour of $B_{z0}$.}
        \label{fig:magf}
\end{minipage}\hfill
\end{figure*}

\begin{figure*}[h!]
\begin{minipage}{\columnwidth}
        \resizebox{\hsize}{!}{  \includegraphics{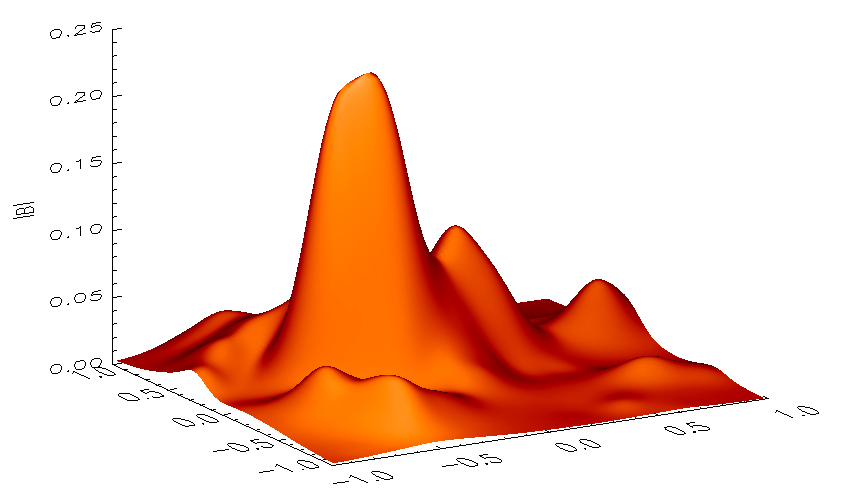}}
        \caption{Shaded surface diagram of the magnetic field strength at $z=0$ for the potential field generated using solar magnetogram data and a 
        standard input region.}
        \label{fig:flared2}
\end{minipage}\hfill
\begin{minipage}{\columnwidth}
        \resizebox{\hsize}{!}{  \includegraphics{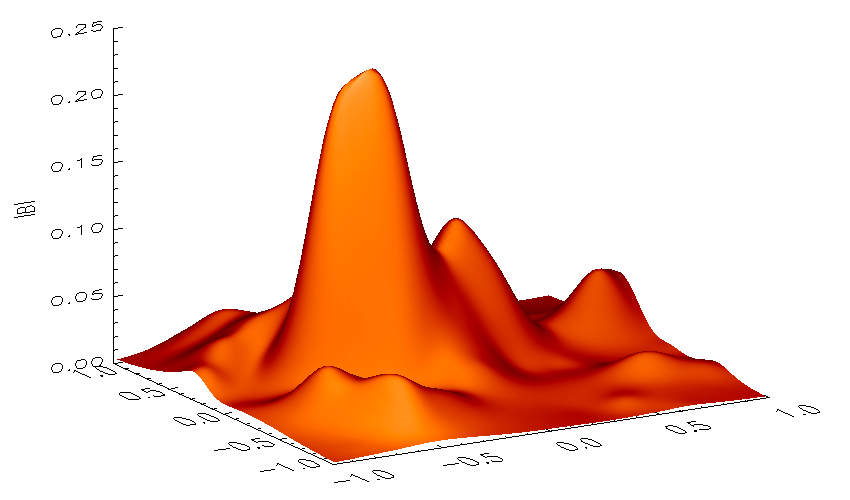}}
        \caption{Shaded surface diagram of the magnetic field strength at $z=0$ for the potential field generated using solar magnetogram data and 
        an extended input region.}
        \label{fig:smooth2}
\end{minipage}\hfill
\end{figure*}

\section{Solar magnetogram data}
\subsection{Method}
Having determined the preferred methodology for our potential calculator and compared this to an analytical solution, we conclude our analysis by considering the magnetic field generated from solar magnetogram data. The data used was taken from line-of-sight magnetogram data collected by the Helioseismic and Magnetic Imager (HMI) instrument on the Solar Dynamics Observatory (SDO) and accessed through the Joint Science Operations Center (JSOC) on-line database. The data was taken from an active region close to the centre of the solar disc between 14:02 and 14:56 on 31/08/2003. As the data was taken from a small area in the centre disc, it was taken to approximately represent the normal magnetic field data in that area. The data was normalised so that the maximum field strength was one and therefore the results are comparable with those from earlier sections. 

A contour plot of the magnetogram data is given in \cref{fig:bz0} and the potential field produced using this magnetogram data is represented by a field line diagram shown in \cref{fig:magf}. Using this data potential fields were produced using each of the three methods of numerical differentiation described in \cref{section5} for a standard input region, as described in \cref{section6}, and using our preferred Method C for numerical differentiation for an extended input region. 

\subsection{Results}
The potential fields produced were run in Lare3d and analysed in the same way as for the previous magnetic fields. The potential field produced using Method A for numerical differentiation, which ignores the effect of the staggered grid, would not run for even $5\tau_A$ and had initial maximum currents an order of magnitude larger than any of the other potential fields. The results for the analysis of the remaining potential fields are given in \cref{tab:6}. 

\begin{table}[!h]
\caption{Value of each metric for each of the potential fields produced using the solar magnetogram data.}
\label{tab:6}
\centering
\begin{tabular}{ |p{0.2\columnwidth}||p{0.2\columnwidth}|p{0.2\columnwidth}|p{0.2\columnwidth}| }
 \hline
 & Method B (standard input) & Method C (standard input) & Method C (extended input) \\
 \hline
 MAJ$_0$                & $2.36 \cdot 10^{-2}$ & $1.23 \cdot 10^{-2}$ & $1.23 \cdot 10^{-2}$\\
 J$_{avg0}$             & $1.55 \cdot 10^{-5}$ & $7.88 \cdot 10^{-6}$ & $7.90 \cdot 10^{-6}$\\
 MAJ$_f$                & $4.92 \cdot 10^{-2}$ & $4.90 \cdot 10^{-2}$ & $4.90 \cdot 10^{-2}$\\
 J$_{avgf}$                     & $4.45 \cdot 10^{-5}$ & $3.98 \cdot 10^{-5}$ & $4.03 \cdot 10^{-5}$\\
 MAV                    & $1.95 \cdot 10^{-3}$ & $1.81 \cdot 10^{-3}$ & $1.34 \cdot 10^{-3}$\\
 $\Delta B_{max}$       & $4.14 \cdot 10^{-3}$ & $1.91 \cdot 10^{-3}$ & $1.91 \cdot 10^{-3}$\\
 \hline
\end{tabular}
\end{table}

Comparing the results for using Method B or Method C for numerical differentiation, we can see that, consistent with our earlier results, Method C produces lower initial currents. Although over time the fields approach similar current values, the field produced using Method C maintains lower currents, changes less over time, and results in a more static equilibrium. This result supports our conclusion that Method C is the more accurate method in addition to being computationally quicker and requiring less ghost cells.

Comparing now the results for using either a standard or extended input region (both using Method C for differentiation) we see a surprising result. The maximum current is identical in both cases and the average current is marginally lower when the standard input region is used. Although the maximum velocity at $100\tau_A$ is lower when the extended input region is used this difference is less than an order of magnitude. The reason is that using an extended input region has little effect, which can clearly be seen by comparing the magnetic field strength for the potential field generated with the standard input region and extended input region. The field strength at the photosphere is shown for each of these fields in \cref{fig:flared2,fig:smooth2}. It can clearly be seen that in contrast to \cref{fig:smooth} in \cref{section6} the field strength in \cref{fig:smooth2} is much smaller and flatter at the boundaries so assuming $B_{z0}=0$ outside of the domain has much less of an effect.

Although extending the input region in general improves the accuracy of the potential field solution, these results indicate that the benefits of extending the input region size are sometimes only marginal, depending on the input field $B_{z0}$. It must be remembered that to achieve the slight reduction in maximum velocity at $100\tau_A$, the input region had to be extended to roughly nine times. This means that nine times more initial data is required and it also means that the runtime will increase ninefold.  
\section{Conclusion}
\label{section8}
A potential field calculator has been designed to accurately calculate a potential magnetic field over a domain given the normal field $B_{z0}$ at the photospheric boundary. The design of the code was aimed at increasing accuracy by taking into consideration: the calculation of magnetic potential $\phi$, the differentiation of magnetic potential $\phi$, and the size of input region for $B_{z0}$ as described in \cref{section3,section5,section6} of this paper, respectively. 

The potential field calculator uses \cref{eq:8} to calculate the magnetic potential. The potential is then differentiated using a novel central difference formula \cref{eq:13}, derived in \ref{App1}, to calculate the magnetic field components. It has been shown that ignoring the staggered magnetic fields during differentiation of the magnetic potential results in an unstable potential field and that, whilst simple interpolation can be effective, the chosen method of numerical differentiation is ultimately more accurate, computationally quicker, and requires less ghost cells to perform. Further it has been shown in \cref{section7} that for analytically specified input data, this method can produce potential fields that are correct to within the numerical diffusion of the simulation grid.

The size of the input region, over which the normal magnetic field at the photosphere $B_{z0}$ is specified, has been shown to effect the accuracy of the potential fields generated. It has been shown that in some cases extending the input region dramatically improves the accuracy of the potential field solution whilst in other cases the benefits of extending the input region size are only marginal. In all cases however extending the input region requires more input data to be specified and increases the runtime of the potential calculator. Depending on the improvement to the potential field solution, the availability of photospheric data, and the desired runtime of the calculator, it may or may not be beneficial to use an extended input region. The potential field calculator therefore leaves the extension of the input region as an option to the user.

It should be noted that the equilibria produced using the potential fields generated may be further improved within an MHD solver by applying a phase of numerical relaxation. In Lare3d this is often done by artificially increasing the resistivity and allowing the system to relax before a simulation is run properly.

Whilst the Green's function methods used in this work have been available for a considerable length of time \citep{Schmidt,Semel,sakurai}, it is hoped that this accurate and simple implementation will be of considerable utility to those working with potential fields on a staggered grid. The potential calculator code is written in FORTRAN 90 and has been designed to produce initial magnetic field inputs for Lare3d but will work with any MHD solver that uses a staggered grid for the magnetic field components.  The program can be parallelised to run quickly over multiple computing cores. This code and supporting documentation are provided alongside this paper. \\

\subsection{Acknowledgements}
Callum Boocock would like to thank UK STFC DISCnet for financial support of his PhD studentship.
This research utilised Queen Mary's Apocrita HPC facility, supported by QMUL Research-IT \citep{Apoc}.
\bibliography{PotentialCalc}{}
\bibliographystyle{aa}

\newpage
\onecolumn
 \appendix
 \setcounter{figure}{0}
 \section{Derivation of formula for numerical differentiation}
 \label{App1}
Our potential field calculator calculates the magnetic field components by taking derivates of the magnetic potential according to \cref{eq:5} of the main text. The method of numerical differentiation used is a central difference formula of order $\mathcal{O}(h^4)$ that has been adapted to work on a staggered grid. In this Appendix we derive this formula.

\begin{figure}[!h]
\centering
\captionsetup{width=0.6\linewidth}
\includegraphics[width=0.6\linewidth]{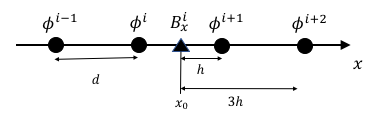}
\caption{One-dimensional representation of a small section around $B^{i}_x$ at a point $x_0$ in the $x$ direction showing the position of $\phi$ values relative to $x_0$ in the $x$ direction.}
\label{fig:methodc}
\end{figure}

We take the calculation of $B_x(i,j,k)$ as an example, the indices $j$ and $k$ are taken to be fixed and $B_x(i,j,k)$ is denoted simply as $B^{i}_x$. We now consider the position of  $B^{i}_x$ relative to the points, where $\phi$ is defined along the x-axis. If we define $d$ as the grid spacing and $h$ as half the grid spacing, then for $B^{i}_x$ defined at a point $x_0$, $\phi$ is defined at distances $h$ and $3h$ either side of $x_0$. This is illustrated in \cref{fig:methodc}. Now we calculate the derivative of $\phi$ with respect to $x$ at the point $x_0$ that is $\phi'(x_0)$. We begin by taking the Taylor expansions of $\phi^{i-1},\phi^{i},\phi^{i+1}$, and $\phi^{i+2}$ about $x$ these are given below in \cref{eq:A1,eq:A2,eq:A3,eq:A4} as follows:

\begin{equation}\label{eq:A1}
\phi^{i-1}  = \phi(x_0-3h) = \phi(x_0) - 3h\phi'(x_0) +\frac{9h^2}{2}\phi''(x_0)-\frac{27h^3}{6}\phi'''(x_0)+\mathcal{O}(h^4),
\end{equation}
\begin{equation}\label{eq:A2}
\phi^{i}     = \phi(x_0-h)    = \phi(x_0) - h\phi'(x_0) +\frac{h^2}{2}\phi''(x_0)-\frac{h^3}{6}\phi'''(x_0)+\mathcal{O}(h^4),
\end{equation}
\begin{equation}\label{eq:A3}
\phi^{i+1} = \phi(x_0+h)  = \phi(x_0) + h\phi'(x_0) +\frac{h^2}{2}\phi''(x_0)+\frac{h^3}{6}\phi'''(x_0)+\mathcal{O}(h^4),
\end{equation}
\begin{equation}\label{eq:A4}
\phi^{i+2} = \phi(x_0+3h) = \phi(x_0) + 3h\phi'(x_0) +\frac{9h^2}{2}\phi''(x_0)+\frac{27h^3}{6}\phi'''(x_0)+\mathcal{O}(h^4).\end{equation}

First we eliminate the even powers of $h$ by taking $\phi^{i+2}-\phi^{i-1}$ and $\phi^{i+1}-\phi^{i}$ as shown below in \cref{eq:A5,eq:A6}:

\begin{equation}\label{eq:A5}
\phi^{i+2}-\phi^{i-1}  = 6h\phi'(x_0) +9h^3\phi'''(x_0)+\mathcal{O}(h^4),
\end{equation}
\begin{equation}\label{eq:A6}
\phi^{i+1}-\phi^{i}  =  2h\phi'(x_0) +\frac{h^3}{3}\phi'''(x_0)+\mathcal{O}(h^4).\end{equation}

We then need to remove the third order term by taking $27(\phi^{i+1}-\phi^{i})-(\phi^{i+2}-\phi^{i-1})$ as shown below in \cref{eq:A7}:

\begin{equation}\label{eq:A7}
27(\phi^{i+1}-\phi^{i})-(\phi^{i+2}-\phi^{i-1}) = 48h\phi'(x_0)+\mathcal{O}(h^4).\end{equation}

Rearranging this equation we have the central difference formula,

\begin{equation}\label{eq:A8}
B^{i}_{x}= \phi'(x_0)= \frac{\phi^{i-1}-27\phi^{i}+27\phi^{i+1}-\phi^{i+2}}{48h}.\end{equation}

\newpage
 \section{Potential calculator - BooTsik.f90}
 \label{App2}
 This Appendix provides the FORTRAN 90 code for the potential calculator, BooTsik.f90, which can also be found at \url{https://github.com/calboo/Potential\_Field\_Calculator}. The user must set the domain size, nx, ny, nz; the grid spacing, d; the size of the input region, nx0, nx1, ny0, ny1, and must specify the normal field $B_{z0}$ at the photosphere. There are preset options for the input fields $B_{z0}$ used in this paper and for importing an input field from an unformatted data file that can be used simply by uncommenting them. The code is given below:
 \begin{scriptsize}
 \begin{verbatim} 
! BooTsik.f90
! A FORTRAN 90 code designed to calculate a potential magnetic field 
! by extrapolating the normal field given at the base of the domain.
! 
! Authors: Callum Boocock and David Tsiklauri
! Institution: QMUL
! Email: c.boocock@qmul.ac.uk
! Date : 3rd November 2018
!
! Copyright (C) 2018 
!
! This program is free software: you can redistribute it and/or modify
! it under the terms of the GNU General Public License as published by
! the Free Software Foundation, either version 3 of the License, or
! any later version.
!
! This program is distributed in the hope that it will be useful,
! but WITHOUT ANY WARRANTY; without even the implied warranty of
! MERCHANTABILITY or FITNESS FOR A PARTICULAR PURPOSE.  See the
! GNU General Public License for more details:
! <http://www.gnu.org/licenses/>.
!

MODULE constants
IMPLICIT NONE

! Dimension of domain in which to calculate 3D potential field.
INTEGER, PARAMETER :: nx = 100, ny = 100, nz = 100
!
! Set parameter for grid spacing.
DOUBLE PRECISION, PARAMETER :: d=0.01
!
! Set size of input region
! Uncomment 1st line for standard, 2nd line for extended
!DOUBLE PRECISION, PARAMETER :: nx0=-1, nx1=nx+2, ny0=-1, ny1=ny+2
!DOUBLE PRECISION, PARAMETER :: nx0=-nx-3, nx1=2*nx+4, ny0=-ny-3, ny1=2*ny+4
!
! Parameters for defining normal field Bz0 at base of domain z=0.
DOUBLE PRECISION, PARAMETER :: x01=0.0, y01=0.0, x02=0.0, y02=0.5, sigma=1.0, amp=1.0
!
! Value for pi
DOUBLE PRECISION, PARAMETER :: pi = 3.14159265358979323
!
! Arrrays and indices
DOUBLE PRECISION, DIMENSION(-3:nx+4,-3:ny+4,-3:nz+4) :: phi
DOUBLE PRECISION, DIMENSION(-2:nx+2,-1:ny+2,-1:nz+2) :: bx
DOUBLE PRECISION, DIMENSION(-1:nx+2,-2:ny+2,-1:nz+2) :: by
DOUBLE PRECISION, DIMENSION(-1:nx+2,-1:ny+2,-2:nz+2) :: bz 
REAL, DIMENSION(nx0:nx1,ny0:ny1) :: bz0
INTEGER :: i,j,k,i1,j1

END MODULE constants

PROGRAM potential
USE constants 
IMPLICIT NONE
phi=0.0

!!!!!!!!!!!!!!!!!!!!!!!!!!!!!!!!!!!!!!!!!!!!!!!!!!!!!!!!!!!!!!!!!!!!!!!!!!!!!!!!!!!!!!!!!!!!!
! Below are three options for the Bz0 field that needs to be specified at the lower boundary.
! The options are a)Bipole loop, b)Gaussian pole, c)Input from data file. For options (a) and 
! (b) the parameters can be set at the top of this code. Uncomment whichever Bz0 profile you
! want to use or alternatively specify your own Bz0 field.

!Bipole loop
!do j=ny0,ny1
!do i=nx0,nx1
!bz0(i,j)= amp*((1.0/((((real(i)-(nx/2.0))/(nx/2.0))-x01)**2+&
!                     (((real(j)-(ny/2.0))/(ny/2.0))-y01)**2+1.0**2)**(3.0/2.0))&
!              -(1.0/((((real(i)-(nx/2.0))/(nx/2.0))-x02)**2+&
!                     (((real(j)-(ny/2.0))/(ny/2.0))-y02)**2+1.0**2)**(3.0/2.0)))
!end do
!end do

!Gaussian pole
!do j=ny0,ny1
!do i=nx0,nx1
!bz0(i,j)=amp*exp(-((((real(i)-(nx/2.0))/(nx/2.0))-x01)**2.0+ &
!                   (((real(j)-(ny/2.0))/(ny/2.0))-y01)**2.0+1.0**2.0)/(2.0*sigma**2))
!end do
!end do

!Read in Bz0 from unformatted data file
!Make sure that the size of the imported array is equal to the size of bz0 as defined in constants. 
!OPEN(unit=12, FORM = 'UNFORMATTED', STATUS='OLD', ACTION='READ', FILE = 'bz0.dat')
!READ(12) bz0
!CLOSE(12)

!!!!!!!!!!!!!!!!!!!!!!!!!!!!!!!!!!!!!!!!!!!!!!!!!!!!!!!!!!!!!!!!!!!!!!!!!!!!!!!!!!!!!!!!!!!!!
! This loop calculates the magnetic potential using a Modified Green's Function method
! 
do k=-3,nz+4
do j=-3,ny+4
do i=-3,nx+4
do i1=nx0,nx1
do j1=ny0,ny1
phi(i,j,k)=phi(i,j,k)+bz0(i1,j1)*d/(2.0*pi*sqrt( (i-i1)**2+(j-j1)**2+ (k+4+1/sqrt(2.0*pi))**2 ) )
end do
end do
end do
end do
print*,'Calculating Phi at HIGHT =',k+4,' out of ',nz+8
end do

!!!!!!!!!!!!!!!!!!!!!!!!!!!!!!!!!!!!!!!!!!!!!!!!!!!!!!!!!!!!!!!!!!!!!!!!!!!!!!!!!!!!!!!!!!!!!
! This section calculates the magnetic field components by numerical differentiation
! of the magnetic potential.
! 
bx=0.0
by=0.0
bz=0.0
do k=-1,nz+2
do j=-1,ny+2
do i=-2,nx+2
bx(i,j,k)=-(phi(i-1,j,k)-27.0*phi(i,j,k)+27.0*phi(i+1,j,k)-phi(i+2,j,k))/(24.0*d)
end do
end do
print*,'Now calculating Bx at HEIGHT =',k+2,' out of ',nz+4
end do
do k=-1,nz+2
do j=-2,ny+2
do i=-1,nx+2
by(i,j,k)=-(phi(i,j-1,k)-27.0*phi(i,j,k)+27.0*phi(i,j+1,k)-phi(i,j+2,k))/(24.0*d)
end do
end do
print*,'Now calculating By at HEIGHT =',k+2,' out of ',nz+4
end do
do k=-2,nz+2
do j=-1,ny+2
do i=-1,nx+2
bz(i,j,k)=-(phi(i,j,k-1)-27.0*phi(i,j,k)+27.0*phi(i,j,k+1)-phi(i,j,k+2))/(24.0*d)
end do
end do
print*,'Now calculating Bz at HEIGHT =',k+2,' out of ',nz+4
end do

!!!!!!!!!!!!!!!!!!!!!!!!!!!!!!!!!!!!!!!!!!!!!!!!!!!!!!!!!!!!!!!!!!!!!!!!!!!!!!!!!!!!!!!!!!!!!
! This section writes the magnetic field components to three files
! bx.dat, by.dat and bz.dat
! 
OPEN(unit=66, FORM = 'UNFORMATTED', FILE = 'bx.dat')
WRITE(66) bx
CLOSE(66)
OPEN(unit=66, FORM = 'UNFORMATTED', FILE = 'by.dat')
WRITE(66) by
CLOSE(66)
OPEN(unit=66, FORM = 'UNFORMATTED', FILE = 'bz.dat')
WRITE(66) bz
CLOSE(66)
END PROGRAM  
 \end{verbatim}
\end{scriptsize}
This code can be run in parallel on a HPC cluster. To achieve this the code can be auto-parallelised, using for example, an intel Fortran compiler. Using an Intel Fortran compiler the following line compiles the code in parallel ready to be submitted as a job on a HPC node. This significantly improves the runtime, depending on the number of cores used. The line is written as

 \begin{scriptsize}
 \begin{verbatim}
 ifort -o potential -O3 -parallel potential.f90 -xHost.
 \end{verbatim}
 \end{scriptsize}

\newpage
 \setcounter{figure}{0}
\section{IDL visualisation tool - Boobox.pro}
\label{App3}
This appendix provides the IDL script for a visualisation tool, Boobox.pro, for visualising magnetic fields either from unstructured data files or from data snapshots from Lare3d. The fields are visualised as field lines in a 3D box with the base of the box contoured according to the strength and direction of Bz at that boundary. The script was used to generate the field line images in this paper. 

\begin{figure}[h!]
\begin{minipage}{0.49\columnwidth}
        \resizebox{\hsize}{!}{  \includegraphics{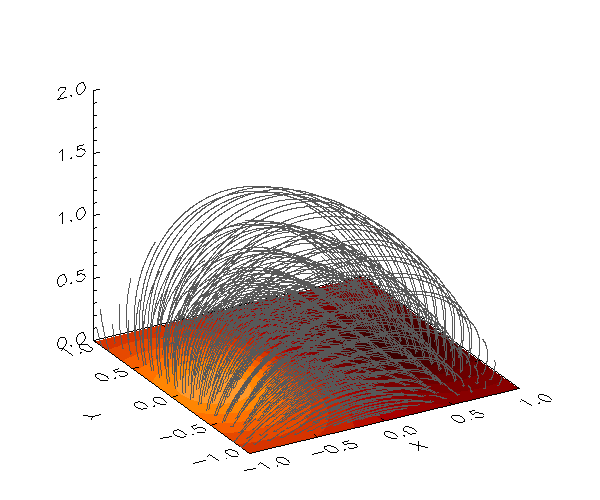}}
        \caption{Field line diagram of a potential bipole field visualised with Boobox.pro with GRID=1. We can clearly see the magnetic field lines 
        connected to each point at the photosphere.}
        \label{fig:grid1}
\end{minipage}\hfill
\begin{minipage}{0.49\columnwidth}
        \resizebox{\hsize}{!}{  \includegraphics{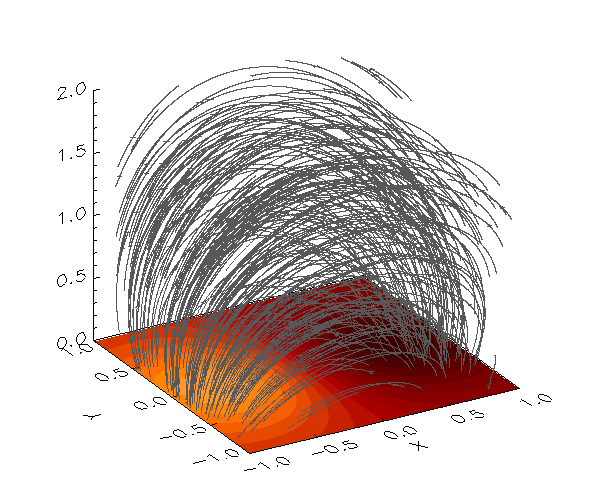}}
        \caption{Field line diagram of a potential bipole field visualised with Boobox.pro with GRID=0. We can clearly see that the field strength is 
        greater closer to the poles at the photosphere.}
        \label{fig:grid0}
\end{minipage}\hfill
\end{figure}

To use the tool the user must specify the domain dimensions, nx, ny, nz; the domain extents for the tick labels and the tick label format; the angles at which to visualise the domain, ax, az, and must finally specify a value for the variable GRID. If GRID is set to the integer value 1 then the field lines are drawn from starting points uniformly distributed across the photospheric surface at the base of the domain. If GRID is set to any other value then the field lines are drawn from starting points randomly distributed throughout the domain, producing a plot where the density of the field lines corresponds roughly to the field strength. Examples of plots produced for a bipole loop using Boobox.pro with GRID set to 1 and 0 are given in \cref{fig:grid0,fig:grid1}. 

If the tool is being used with unstructured data files of the magnetic field components bx.dat, by.dat, and bz.dat, then the appropriate block must be uncommented and the tool can be run in IDL with \texttt{.r boobox}. If the tool is being used to visualise data from a Lare3d data snapshot, ds, then the first line \texttt{pro boobox, ds,} and the appropriate block must be uncommented and the tool can be run in IDL with \texttt{boobox, ds}, where ds is the data snapshot. The IDL script for Boobox.pro is given below:

 \begin{scriptsize}
 \begin{verbatim}
; Boobox.pro
;
; An IDL script for visualising magnetic fields either from  unstructured data files or from data snapshots from Lare3d. 
; The fields are visualised as field lines in a 3D box with the base of the box contoured according to the strength and dirction of Bz at
; that boundary.
; 
; Author: Callum Boocock
; Institution: QMUL
; Email: c.boocock@qmul.ac.uk
; Date : 3rd November 2018
;
; Copyright (C) 2018 
;
; This program is free software: you can redistribute it and/or modify
; it under the terms of the GNU General Public License as published by
; the Free Software Foundation, either version 3 of the License, or
; any later version.
;
; This program is distributed in the hope that it will be useful,
; but WITHOUT ANY WARRANTY; without even the implied warranty of
; MERCHANTABILITY or FITNESS FOR A PARTICULAR PURPOSE.  See the
; GNU General Public License for more details:
; <http://www.gnu.org/licenses/>.
;
; Uncomment the following line only if using with data snapshot ds from Lare3d
; pro boobox, ds

; Domain dimensions
nx=100.000
ny=100.000
nz=100.000
; Extents of domain, for tick labels
xmin=-1.0
xmax= 1.0
ymin=-1.0
ymax= 1.0
zmin= 0.0
zmax= 2.0
; Format of tick labels
tikform = '(f5.1)'
; Viewing angles for visualisation
ax=30
az=30
; Arrangement of field lines
; 1 -               starting points distributed uniformly at the photosphere
; Any other value - starting points randomly distributed throughout domain 
GRID = 0

; Uncomment this block for use with .dat files
;
;bx0=dblarr(nx+5,ny+4,nz+4)
;by0=dblarr(nx+4,ny+5,nz+4)
;bz0=dblarr(nx+4,ny+4,nz+5)
;x=dblarr(nx+4)
;y=dblarr(ny+4)
;z=dblarr(nz+4)
;OPENR, 1, 'bx.dat',  /F77_UNFORMATTED
;READU, 1,bx0  
;CLOSE, 1  
;OPENR, 1, 'by.dat',  /F77_UNFORMATTED
;READU, 1,by0
;CLOSE, 1  
;OPENR, 1, 'bz.dat',  /F77_UNFORMATTED
;READU, 1,bz0
;CLOSE, 1 
;for i=0,nx+3 do x(i)=i
;for i=0,ny+3 do y(i)=i
;for i=0,nz+3 do z(i)=i
;bx0=bx0(2:nx+3,2:ny+2,2:nz+2)
;by0=by0(2:nx+2,2:ny+3,2:nz+2)
;bz0=bz0(2:nx+2,2:ny+2,2:nz+3)

; Uncomment this block for use with data snapshot ds from Lare3d
;
;bx0 = ds.bx
;by0 = ds.by
;bz0 = ds.bz
;x = ds.x
;y = ds.y
;z = ds.z

; Calculate and format the tick labels
x2=(xmin+xmax)/2.0
x1=(xmin+x2)/2.0
x3=(x2+xmax)/2.0
y2=(ymin+ymax)/2.0
y1=(ymin+y2)/2.0
y3=(y2+ymax)/2.0
z2=(zmin+zmax)/2.0
z1=(zmin+z2)/2.0
z3=(z2+zmax)/2.0
xmin = string(xmin,FORMAT=tikform)
x2 = string(x2,FORMAT=tikform)
x1 = string(x1,FORMAT=tikform)
x3 = string(x3,FORMAT=tikform)
xmax = string(xmax,FORMAT=tikform)
ymin = string(ymin,FORMAT=tikform)
y2 = string(y2,FORMAT=tikform)
y1 = string(y1,FORMAT=tikform)
y3 = string(y3,FORMAT=tikform)
ymax = string(ymax,FORMAT=tikform)
zmin = string(zmin,FORMAT=tikform)
z2 = string(z2,FORMAT=tikform)
z1 = string(z1,FORMAT=tikform)
z3 = string(z3,FORMAT=tikform)
zmax = string(zmax,FORMAT=tikform)
xt=[strtrim(xmin,2),strtrim(x1,2),strtrim(x2,2),strtrim(x3,2),strtrim(xmax,2)]
yt=[strtrim(ymin,2),strtrim(y1,2),strtrim(y2,2),strtrim(y3,2),strtrim(ymax,2)]
zt=[strtrim(zmin,2),strtrim(z1,2),strtrim(z2,2),strtrim(z3,2),strtrim(zmax,2)]

; Set colours
TVLCT, 255, 255, 255, 254 ; White color
TVLCT, 0, 0, 0, 253       ; Black color
TVLCT, 88, 88, 88, 252    ; Line color
!P.Color = 253
!P.Background = 254

; Clear window
WINDOW, 1, XSIZE=600, YSIZE=500, TITLE='Boobox'
ERASE

; Set up the 3D scaling system:
SCALE3, xr=[0,nx], yr=[0,ny], zr=[0,nz],ax=ax,az=az

; Contour the base:
CONTOUR,bz0(*,*,0),/fill,nlevels=25,/t3d,/noerase,zvalue=0.0,/noclip,$
XSTYLE=1,YSTYLE=1,XRANGE=[0,nx],YRANGE=[0,ny],CHARSIZE=4,$
XTITLE='X',YTITLE='Y',POS=[0.1,0.1,nx,ny],$
XTICKS=4,XTICKNAME=xt,$
YTICKS=4,YTICKNAME=yt,$
ZTICKS=4,ZTICKNAME=zt

; Set the 3D coordinate space with axes.
 SURFACE, [[0,nx],[0,ny]], /NODATA,/SAVE,/noerase,/t3d,$
XSTYLE=1,YSTYLE=1,/noclip,XRANGE=[0,nx],YRANGE=[0,ny],ZRANGE=[0,nz],$
CHARSIZE=4,POS=[0.1,0.1,nx,ny],$
XTICKS=4,XTICKNAME=xt,$
YTICKS=4,YTICKNAME=yt,$
ZTICKS=4,ZTICKNAME=zt

; Plot the vector field
!P.Color = 252
IF (GRID eq 1) THEN BEGIN
k=0
n=0
sx=FLTARR(1000)
sy=FLTARR(1000)
sz=FLTARR(1000)
for j=0,ny-1,(ny/20.0) do begin
   for i=0,nx-1,(nx/20.0) do begin
      sx(n)=i
      sy(n)=j
      sz(n)=k
      n=n+1
   endfor
endfor
ENDIF ELSE BEGIN
seed=12345678
sx = FIX(nx* RANDOMU(seed,250))
sy = FIX(ny* RANDOMU(seed,250))
sz = FIX(nz* RANDOMU(seed,250))
ENDELSE
SCALE3, zr=[0,nz],ax=ax,az=az
FLOW3,  bx0, by0, bz0,ARROWSIZE=0.01,LEN=2.0,NSTEPS=9999,sx=sx,sy=sy,sz=sz
FLOW3, -bx0,-by0,-bz0,ARROWSIZE=0.01,LEN=2.0,NSTEPS=9999,sx=sx,sy=sy,sz=sz
!P.Color = 253

; Uncomment to produce a png file
;write_png, 'Fields.png',tvrd(/true)  

end
\end{verbatim}
 \end{scriptsize}
 
\newpage 
\section{Importing potential field data into Lare3d}
\label{App4}
This appendix explains how the potential field data produced by BooTsik.f90 and stored in three files, bx.dat, by.dat, and bz.dat was imported into Lare3d and used as the initial magnetic field configuration. Firstly the data files  bx.dat, by.dat, and bz.dat were copied into the same directory that Lare3d was run from. Changes were then made to four of the source files for Lare3d.
\\ \\
The first source file to be edited was shared\_ data.F90, which can be found through the path src/core/shared\_ data.F90. The following lines were added to the module \texttt{shared\_data} to create allocatable arrays:

\begin{scriptsize}
\begin{verbatim}
! Additional allocatable arrays for:
! Storing input magnetic field data
! Storing initial values for simulation variables
! Calculating magnetic fields at cell centres

  REAL(num), DIMENSION(:,:,:), ALLOCATABLE :: bx_init, by_init, bz_init
  REAL(num), DIMENSION(:,:,:), ALLOCATABLE :: rho0, energy0, bx0, by0, bz0
  REAL(num), DIMENSION(:,:,:), ALLOCATABLE :: arrx,arry,arrz
\end{verbatim}
\end{scriptsize}
The second source file to be edited was control.f90, which can be found through the path src/control.f90. The changes that were made to this file are as follows. In the subroutine \texttt{user\_normalisation} the normalisations were all set to  $1.0$. In the subroutine \texttt{control\_variables} the number of gridpoints and the domain size were changed to match our input data, the runtime was set to 100.0 (this is measured in $\tau_A$), the resistivity was set to $5\cdot 10^{-5}$, and the boundary conditions were set to user defined \texttt{BC\_USER}. The following lines were then added to the end of the subroutine \texttt{control\_variables} to read in the potential magnetic field data into allocatable arrays \texttt{bx\_init},\texttt{by\_init} and \texttt{bz\_init}:

\begin{scriptsize}
\begin{verbatim}
    ! Here we allocate the arrays bx_init, by_init and bz_init
    ! We then read the files bx.dat, by.dat and bz.dat
    ! and store the input magnetic fields into these arrays.
    
    ALLOCATE( bx_init(-2:nx_global+2,-1:ny_global+2,-1:nz_global+2))
    ALLOCATE( by_init(-1:nx_global+2,-2:ny_global+2,-1:nz_global+2))
    ALLOCATE( bz_init(-1:nx_global+2,-1:ny_global+2,-2:nz_global+2))

    OPEN(unit=12, FORM = 'UNFORMATTED', STATUS='OLD', ACTION='READ', FILE = 'bx.dat')
    READ(12) bx_init(-2:nx_global+2,-1:ny_global+2,-1:nz_global+2)

    OPEN(unit=12, FORM = 'UNFORMATTED', STATUS='OLD', ACTION='READ', FILE = 'by.dat')
    READ(12) by_init(-1:nx_global+2,-2:ny_global+2,-1:nz_global+2)

    OPEN(unit=12, FORM = 'UNFORMATTED', STATUS='OLD', ACTION='READ', FILE = 'bz.dat')
    READ(12) bz_init(-1:nx_global+2,-1:ny_global+2,-2:nz_global+2)
\end{verbatim}
\end{scriptsize}
Finally in the subroutine \texttt{set\_output\_dumps} the time between snapshots was set to 10.0 (this is measured in $\tau_A$) and the dump masks 17-19 for the currents were set to \texttt{.TRUE.}.
\\ \\
The third source file to be edited was initial\_conditions.f90, which can be found through the path src/initial\_conditions.f90. The file initial\_conditions.f90 sets the initial values of density $\rho$, energy $\varepsilon$, velocity field $\mathbf{v}$, and magnetic field $\mathbf{B}$ for each subdomain after domain decomposition. In this file we replace the subroutine \texttt{set\_initial\_conditions} with the following:

\begin{scriptsize}
\begin{verbatim}
  SUBROUTINE set_initial_conditions

  INTEGER :: ix, iy, iz
  REAL(num) :: beta

! Gravity and beta 

  grav=0.0_num
  beta = 1.0e-8_num 

! Velocities
! Static domain.

  vx = 0.0_num
  vy = 0.0_num
  vz = 0.0_num

! Magnetic Field
! The magnetic field components are read in from the arrays
! bx_init, by_init, bz_init.

  DO ix = -2, nx+2
     DO iy = -1, ny+2
         DO iz = -1, nz+2
            bx(ix,iy,iz)=bx_init(ix+n_global_min(1),iy+n_global_min(2),iz+n_global_min(3))
         END DO
     END DO
  END DO

  DO ix = -1, nx+2
     DO iy = -2, ny+2
         DO iz = -1, nz+2
            by(ix,iy,iz)=by_init(ix+n_global_min(1),iy+n_global_min(2),iz+n_global_min(3))
         END DO
     END DO
  END DO

  DO ix = -1, nx+2
     DO iy = -1, ny+2
         DO iz = -2, nz+2
            bz(ix,iy,iz)=bz_init(ix+n_global_min(1),iy+n_global_min(2),iz+n_global_min(3))
         END DO 
     END DO
  END DO

! Density
! In this case the density has been set equal to the
! square of the magnetic field strength.

  ALLOCATE(arrx(-1:nx+2, -1:ny+2, -1:nz+2))
  ALLOCATE(arry(-1:nx+2, -1:ny+2, -1:nz+2))
  ALLOCATE(arrz(-1:nx+2, -1:ny+2, -1:nz+2))

  DO iy= -1,ny+2 
     DO iz = -1,nz+2 
         DO ix = -1,nx+2 
             arrx(ix,iy,iz)=0.5*bx(ix-1,iy,iz)+bx(ix,iy,iz) 
             arry(ix,iy,iz)=0.5*by(ix,iy-1,iz)+by(ix,iy,iz) 
             arrz(ix,iy,iz)=0.5*bz(ix,iy,iz-1)+bz(ix,iy,iz)
             rho(ix,iy,iz) =(arrx(ix,iy,iz)**2.0)+ &
                            (arry(ix,iy,iz)**2.0)+ &
                            (arrz(ix,iy,iz)**2.0)
          END DO
     END DO
  END DO

! Energy
! The energy has been set such that the pressure will equal beta/2 everywhere.

  energy=0.5_num*(beta*1.0_num) / ((rho) * (gamma-1.0_num))

! Store Initial Values
! Initial values are stored in these arrays.

  ALLOCATE(rho0(-1:nx+2, -1:ny+2, -1:nz+2))
  ALLOCATE(bx0 (-2:nx+2, -1:ny+2, -1:nz+2))
  ALLOCATE(by0 (-1:nx+2, -2:ny+2, -1:nz+2))
  ALLOCATE(bz0 (-1:nx+2, -1:ny+2, -2:nz+2))
  ALLOCATE(energy0(-1:nx+2, -1:ny+2, -1:nz+2))

  bx0 = bx
  by0 = by
  bz0 = bz
  rho0 = rho
  energy0 = energy

  END SUBROUTINE set_initial_conditions
\end{verbatim}
\end{scriptsize}
For each subdomain this subroutine sets gravity and the initial velocity field to zero. The subroutine sets the initial internal energy $\varepsilon$ such that the initial pressure $P$ is $5\cdot 10^{-9}$ across the domain, imports the correct subdivision of the potential magnetic field, and sets the density equal to the square of the magnetic field strength $\rho = B^2$. Finally these initial values are stored in arrays \texttt{rho0,bx0,by0,bz0,energy0}.
\\ \\
 The fourth and final source file to be edited was boundary.f90, which can be found through the path src/initial\_boundary.f90. This boundary conditions in this file were edited so that at every boundary the density $\rho$, internal energy $\varepsilon$, and magnetic field components \texttt{bx,by,bz} were all set to their initial values \texttt{rho0,bx0,by0,bz0,energy0}. An example of this is given below, the example given is for the magnetic fields at minimum x-boundary:
 
 \begin{scriptsize}
\begin{verbatim}
    IF (proc_x_min == MPI_PROC_NULL .AND. xbc_min == BC_USER) THEN
      bx(-1,:,:) = bx0(-1,:,:)
      bx(-2,:,:) = bx0(-2,:,:)
      by( 0,:,:) = by0( 0,:,:)
      by(-1,:,:) = by0(-1,:,:)
      bz( 0,:,:) = bz0( 0,:,:)
      bz(-1,:,:) = bz0(-1,:,:)
    END IF
\end{verbatim}
\end{scriptsize}
Finally the velocities at all boundaries at both full and half timesteps were all set to zero. An example of this is given below, the example given is for the full timestep velocities at minimum x-boundary:

 \begin{scriptsize}
\begin{verbatim}
    IF (proc_x_min == MPI_PROC_NULL .AND. xbc_min == BC_USER) THEN
      vx(-2:0,:,:) = 0.0_num
      vy(-2:0,:,:) = 0.0_num
      vz(-2:0,:,:) = 0.0_num
    END IF
\end{verbatim}
\end{scriptsize}
 
\end{document}